\documentclass[aps,prb,showpacs,twocolumn]{revtex4}
\usepackage{graphicx}
\usepackage{amsmath}
\usepackage{color}

\input{epsf}
\begin{document}

\title{Radiation-induced quantum Fano-type resonances in  the transport of $N$-$P$-$N$ graphene based junctions}

\author{M. V. Fistul$^{1,2}$, and K. B. Efetov$^{1,2}$}
\affiliation{$^{1}$ Theoretische Physik III, Ruhr-Universit\"at
Bochum, D-44801 Bochum, Germany }
\affiliation{$^2$ National University of Science and Technology MISIS, Moscow 119049, Russia
}
\date{\today}

\begin{abstract}
We present a theoretical study of quantum resonances in the ballistic transport  of  graphene based $N$-$P$-$N$ junction subject to an externally applied electromagnetic field (EF). By making use of the  Floquet analysis and the quasi-classical approach we analyze the dynamics of electrons in the presence of time and coordinate dependent potential $U(z,t)$. In the absence of EF the resonant tunneling results in a set of sharp resonances in the dependence of dc conductance $\sigma$ on the gate voltage $V_g$. In irradiated $N$-$P$-$N$ junctions we obtain the Fano-type resonances in the dependence of $\sigma(V_g)$. This \emph{coherent quantum-mechanical phenomenon} is  due to the interplay of two effects: the resonant tunneling through quasi-bound states and the quantum-interference effect in the region between the "resonant points", where the resonant absorption (emission) of photons occurs, and junction interfaces.

\end{abstract}

\maketitle


\section{Introduction}

A great attention has been devoted to theoretical and experimental study of the transport in graphene based nanostructures \cite{Graphene1,Graphene2}. Indeed, such  unusual effects as the Klein tunneling \cite{GrNature,Kltunn1}, the photo-current \cite{PHcurrExp1,PHcurrExp2} and ac-induced ratchet effect   \cite{RatchEffect}, just to name a few, have been predicted and observed. Most of these effects has an origin in a specific gapless two-bands electronic spectrum  in graphene, and therefore, the quantum-mechanical dynamics of quasi-particles well described by Dirac equation.

On other hand, a modern technology allows one to prepare diverse nanostructures, e.g. $N$-$P$ and $N$-$P$-$N$ junctions, quantum point contacts etc., where the ballistic transport is determined by propagation of quasi-particles in specially prepared coordinate dependent potential relief $U(z)$ ($z$ is the longitudinal coordinate perpendicular to the interface). This potential relief can be tuned  by application of gate voltages $V_g$. It has been shown in Ref. \cite{Falko} that the ballistic conductance $\sigma(V_g)$ for $N$-$P$ graphene based junctions is determined by the interface transmission $T$, which in turn, strongly depends on the transverse momentum $p_y$ of propagating quasi-particles, i.e.  $t=\exp\left[-{\pi p_y^2}/{(\hbar \mathrm{v}F)}\right]$,
where $\mathrm{v}$ is the Fermi velocity and $F$ is the slope of potential $U(z)$ close to the interface. This transmission is the Landau-Zener tunneling through the effective gap $\Delta_{eff}~\propto~|p_y|$. Therefore, if the interface prepared in graphene based nanoribbons is smooth enough, the slope $F$ is small and the conductivity can be strongly suppressed.

As we turn to more complex structures, i.e. $N$-$P$-$N$ junctions or quantum point contacts, the quantum-mechanical phenomena of \emph{resonant tunneling} occurs. Indeed, as the transverse momentum of electrons is not small, i.e. $p_y \neq 0$, the  transmission $t$ on both interfaces is small, the quasi-particles are confined in the middle area of the junction, and such a confinement results in a set of quasi-bound states \cite{EfSilv}. These quasi-bound states manifest themselves by resonances (sharp peaks) in the dependence of $\sigma(V_g)$. The width of peaks $\delta \sigma$ is determined by the transmission $t$ as $\delta \sigma~\propto~1/t$.

In the presence of an externally applied radiation the quasi-particles in a two-band material \emph{resonantly} interact with radiation as the condition $\epsilon_c(p)-\epsilon_v(p)=\hbar \omega$ is satisfied. Here, $\epsilon_{c(v)}(p)$ are the conduction (valence) bands and $\omega$ is the frequency of external radiation. In $N$-$P$ junctions this condition is satisfied in a small area, only. Near these "resonant points" the dynamical gap $\Delta_R$ appears in the spectrum of quasi-particles, and the dynamical Landau-Zener tunneling through this gap determines the total transmission through the junction \cite{FEprl,SFEprb}. The dynamical gap $\Delta_R$ shows a great similarity to the Rabi frequency \cite{Rabi,Hanggi}, well-known in physics of coherent quantum-mechanical phenomena.
It was shown in Ref. \cite{quantuminterf} that in irradiated $N$-$P$ junctions the electrons (holes) can be confined in a narrow region between the resonant point and junction interface. Moreover, there are two paths for quasi-particles propagating through the junction, and the interference between these two paths manifests itself by large oscillations of the ballistic photocurrent as a function of the gate voltage $V_g$ or the frequency $\omega$ of the radiation. This \emph{coherent quantum phenomenon} resembles Ramsey quantum beating and Stueckelberg oscillations well-known in atomic physics \cite{Ramsey,Stuck}.

In this paper we consider the ballistic transport in an irradiated $N$-$P$-$N$ junction based on graphene nanoribbons. We focus on quantum resonances in the dependence of dc conductivity $\sigma(V_g)$. We show that these \emph{Fano-type quantum resonances } result from the interplay between the resonant tunneling through quasi-bound states \cite{EfSilv} and ac-induced quantum-interference effect \cite{quantuminterf}. The paper is organized as follows: in Section II we present a model of $N$-$P$-$N$ graphene based junction, and by making use of the rotation wave approximation (RWA) we obtain dynamic spectrum of electrons in irradiated graphene nanostructures. In Section III using the quasi-classical analysis we obtain the transmission coefficient of propagating electrons in $N$-$P$-$N$ junctions, and  the dependence of ballistic dc conductance $\sigma$ on the gate voltage $V_g$ will be calculated by making use of Landauer-B\"{u}ttiker approach adopted to a non-stationary case \cite{SFEprb,quantuminterf,Moskalets}.
The Section IV provides discussion and conclusion.

\section{Model and dynamic spectrum of electrons in irradiated graphene nanostructures}
We consider a graphene based $N$-$P$-$N$ junction in the presence of an applied electromagnetic field (EF). Graphene is a particular two-band material, and therefore, in the absence of EF the quasi-particle spectrum is $\epsilon_{c(v)}=\epsilon_{\pm}=\pm \mathrm{v}\sqrt{p_z^2+p_y^2}$, where $\mathrm{v}$ is the electron velocity, $p_{z(y)}$ are the components of electron momentum in the direction perpendicular (parallel) to the interface. The $N$-$P$-$N$ junction is modeled by smooth potential relief $U(z)$, and an applied electromagnetic field is characterized by vector-potential $\vec{A}=(\vec{E}c/\omega)\cos (\omega t)$, where $E=\sqrt{4\pi S/c}$ is the amplitude of EF, and $S$ is the intensity of EF.  This setup is presented in Fig. 1. The total Hamiltonian of electrons in a two-band material in the presence of both external EF and the coordinate-dependent potential is written as
\begin{equation}
    \hat{H}=\hat{\sigma}_z \epsilon_+(p_z, p_y) + 2\Delta_R\hat{\sigma}_y \cos(\omega t)+U(z),\label{Hamiltonian}
\end{equation}
where the interaction of electrons with EF is determined by so-called  dynamical gap $\Delta_R~\propto~\sqrt{S}$. Notice here, that the dynamical gap $\Delta_R$ depends strongly on the polarization \cite{SFEprb,Mai} and the direction of incidence of applied EF , and it reaches a maximum value for a normal incidence of EF and as the EF is polarized in the direction parallel to the interface. For such conditions the explicit expression for $\Delta_R$ has been obtained in Refs. \cite{FEprl,SFEprb}.

\begin{figure}[tbp]
\includegraphics[width=1.2in,angle=-90]{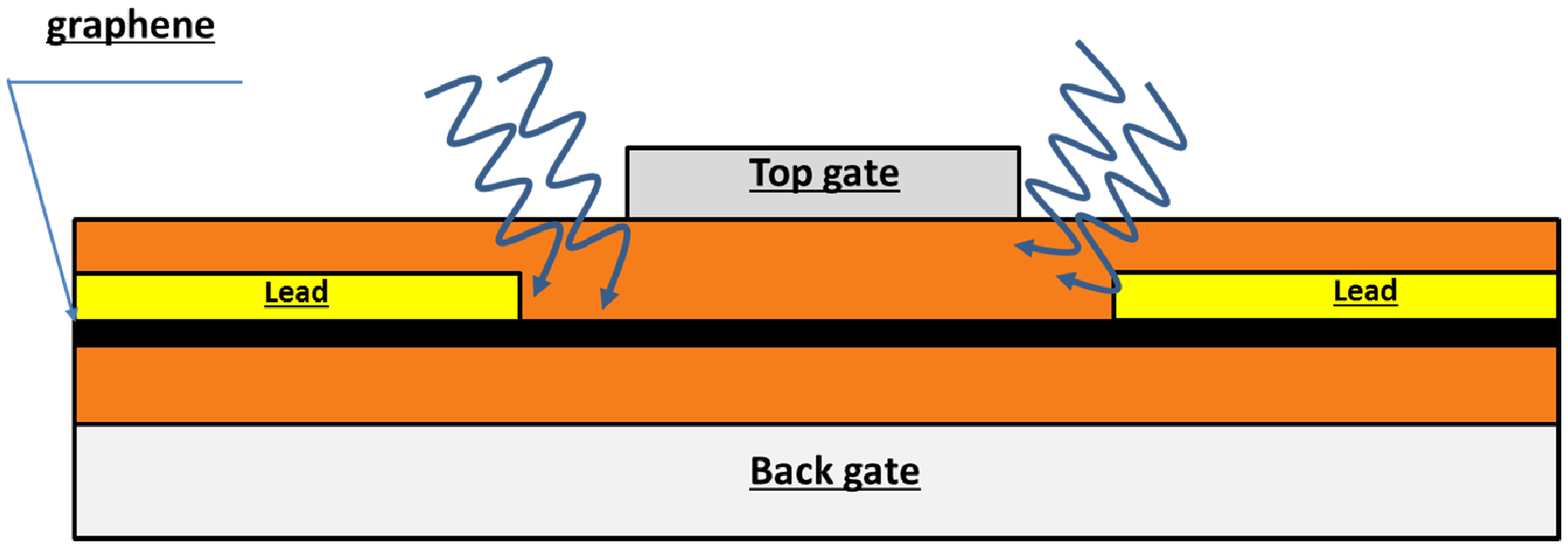}
\includegraphics[width=3in,angle=0]{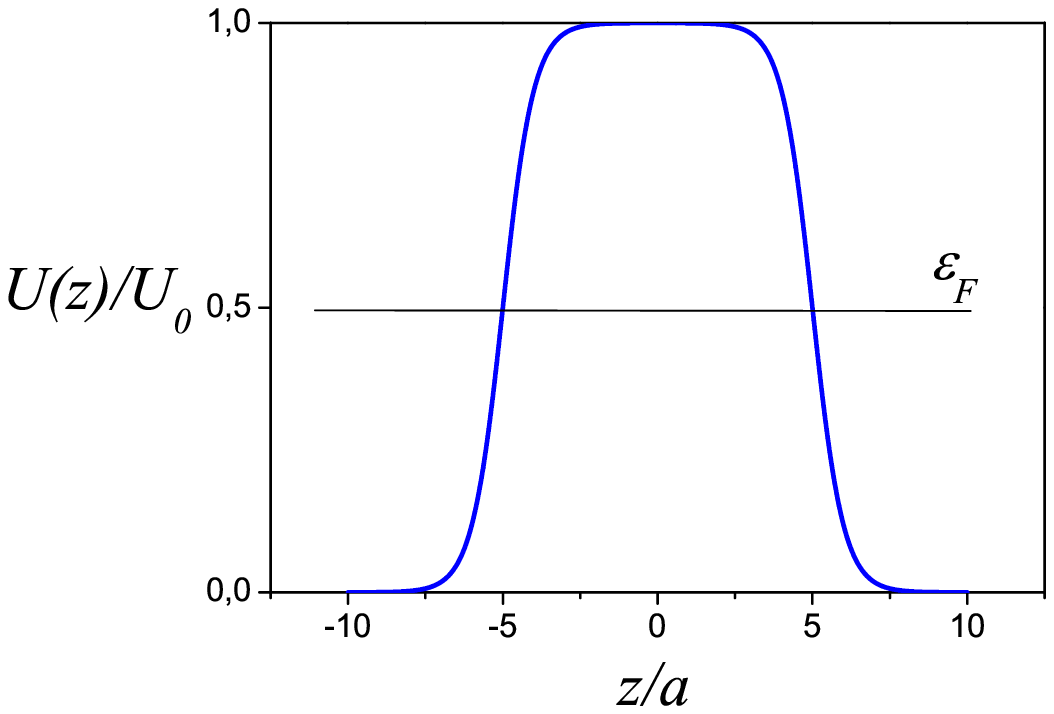}
\caption{Schematic of irradiated graphene $N$-$P$-$N$ junction. The potential relief $U(z)$ is also shown. $\epsilon_F$ is the Fermi energy.
} \label{Schematic}
\end{figure}

In order to obtain the dynamic spectrum of electrons in irradiated graphene we use the unitary transformation of the Hamiltonian as
\begin{equation}
    \hat{H}_{eff}=\hat{U}^+ \hat{H}\hat{U}-i\hbar\hat{U}^+\hat{\dot{U}} ,\label{Hamiltonian-rotation}
\end{equation}
where $\hat{U}=\exp\{-i\omega t \frac{\hat{\sigma}_z }{2} \}$. In the rotation wave approximation (RWA) as the resonant interaction between EF and electrons is taken into account only, we neglect the term $\exp(2i\omega t)$  and  the Hamiltonian  $\hat{H}_{eff}$ does not depend on time explicitly. The Hamiltonian is written as
 \begin{equation}
 \hat{H}_{eff}= \hat{\sigma}_z \left[\epsilon_+(p_z, p_y)-\frac{\hbar \omega}{2} \right] + \Delta_R\hat{\sigma}_x +U(z).\label{dynamical-spectrum}
\end{equation}
As a result we obtain the dynamic spectrum (in the absence of external potential relief, i.e. $U(z)=0$) as
 \begin{equation}
  \tilde{\epsilon}_{n,\pm}=n\hbar \omega \pm \sqrt{(\epsilon_+(\vec{p})-\frac{\hbar \omega}{2})^2+\Delta_R^2}.\label{dynamical-spectrum}
\end{equation}
Thus, one can see that in the presence of externally applied radiation the quasi-particle energy spectrum varies strongly displaying a small energy gap $\Delta_R$ as the resonant condition is satisfied: $\epsilon_+(\vec{p})-\epsilon_-(\vec{p})=\hbar \omega$.

\section{The conductance of irradiated $N$-$P$-$N$ junctions}

Graphene based nanostructures are characterized by specific potential relief $U(z)$. An example of such coordinate-dependent potential for $N$-$P$-$N$ junction is shown in Fig. 1. Moreover, in real nanostructures this potential is smooth enough in order to consider the propagating of electrons in quasi-classical approximation. In this approximation the ballistic transport is determined by phase trajectories of electrons. These phase trajectories written in $\{p_z, z\}$ coordinates are obtained as the solutions of the equations:
 \begin{equation}
 U(z)+\tilde{\epsilon}_{n,\pm}(p_z, p_y)=\epsilon_0\mp \frac{\hbar \omega}{2},\label{equationQC}
\end{equation}
where $\epsilon_0$ is the total energy of propagating electrons, and the component of momentum parallel to the interface, $p_y$, is conserved.

In order to quantitatively analyze different types of phase trajectories we model the potential relief as $U(z)=\frac{U_0}{2}[\tanh{\frac{z+a}{\lambda_\ell}}-\tanh{\frac{z-a}{\lambda_r}}]$, where $2a \gg \lambda_{\ell(r)}$ is the size of $P$-area of the junction, $\lambda_{\ell(r)}$ is the size of left (right) interface region, and $U_0$ can be tuned by external gate voltage.

In the absence of applied EF as $\Delta_R=0$ the phase trajectories are shown in Fig. 2a. In this case
\begin{figure}[tbp]
\includegraphics[width=3.5in,angle=0]{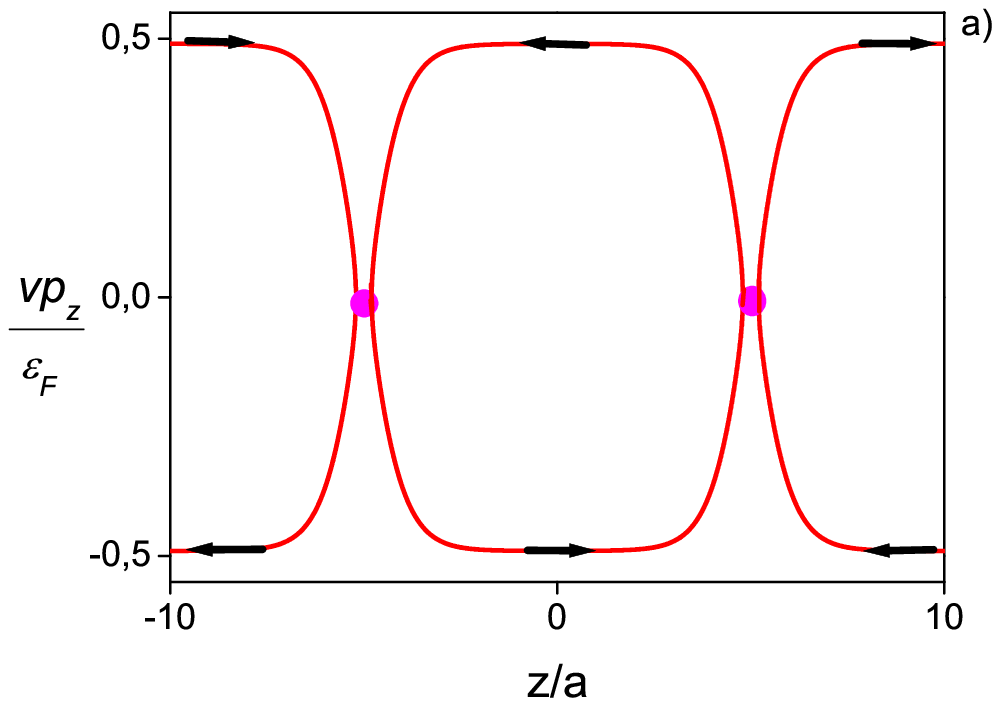}
\includegraphics[width=3.5in,angle=0]{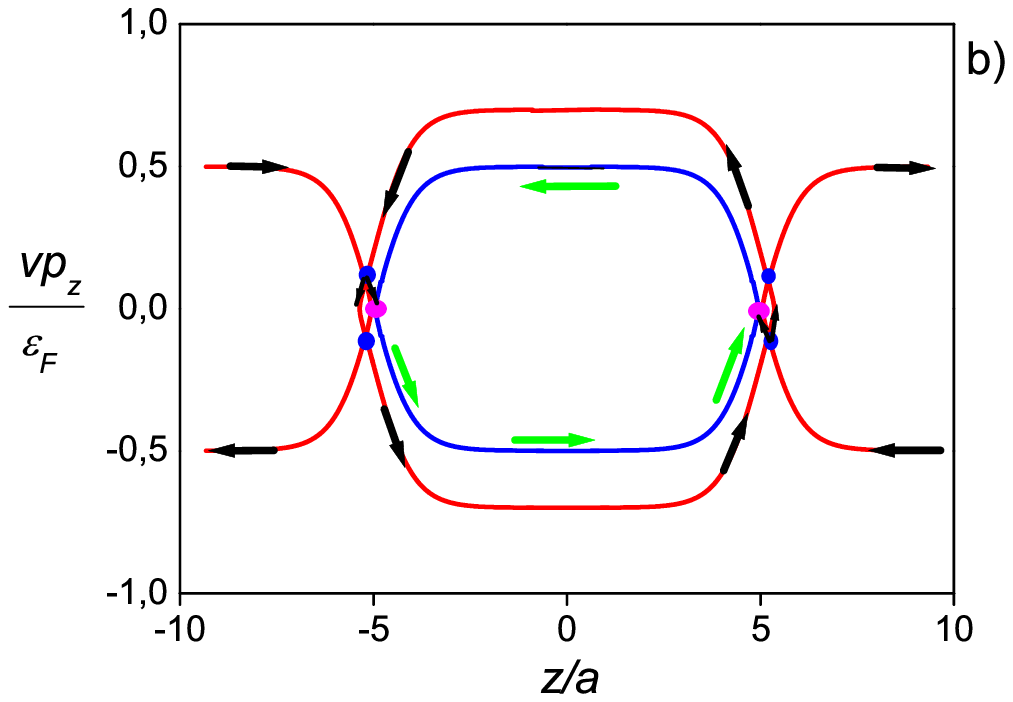}
\caption{The quasi-classical phase trajectories of electrons in $N$-$P$-$N$ junctions:  in the absence (a) and in the presence (b) of EF.  The arrows show the direction of electron motion. The red and blue circles indicate the processes of static and dynamic tunneling, accordingly. The parameters were chosen as $U_0=2\epsilon_F$, $\lambda_{\ell(r)}=0.2 a$, $\mathrm{v}p_y/U_0=0.1$, $\hbar \omega/U_0=0.25$.
} \label{Trajectories}
\end{figure}
the amplitude of electron transmission  from left to right $A_{LR}$ can be obtained by summation of the following series (over all possible pathes)
$$
A_{LR}=a_\ell a_r(1+X+X^2+....),
$$
\begin{equation}
X=b_{\ell}b_{r} \exp\{2i(\Psi+\pi/2)\}\label{transmission},
\end{equation}
where the Landau-Zener tunneling through interfaces is described by amplitudes of transmission $a_{\ell(r)}$ and reflection $b_{\ell(r)}$ on left (right) interfaces. These amplitudes satisfy to the standard condition: $a_{\ell(r)}^2+b_{\ell(r)}^2=1$. For our model the transmission coefficients $t_{\ell(r)}=|a_{\ell(r)}|^2$ are written explicitly: \cite{Falko,SFEprb,quantuminterf}
\begin{equation}
t_{\ell(r)}=\exp \left[-\frac{\mathrm{v}p^2_y\lambda_{\ell(r)}}{\hbar U_0} \right].
\end{equation}
Here, $X$ is the amplitude of quasi-particle propagation along a closed trajectory (see the red (gray) line in the middle of Fig. 2a). Such an amplitude is determined by the total quantum-mechanical phase $\Psi (\epsilon_0)$ of quasi-particles propagating from left turning point (the coordinate $z_\ell$) up to right turning point (the coordinate $z_r$)as
\begin{equation}
\Psi (\epsilon_0)=-(1/\hbar)\int_{z_\ell}^{z_r}p_z(z)dz \label{Phase},
\end{equation}
and  the Stokes phase $\pi/2$ has to be added for each reflection on interfaces [see, Eq. (\ref{transmission})].

Calculating a sum in the Eq. (\ref{transmission}) we obtain $A_{LR}=a_\ell a_r/(1-X)$. In a symmetric junction,  and in the limit of a small Landau-Zener tunneling, i.e $t_{\ell}=t_r =t \ll 1$, the total transmission coefficient through $N$-$P$-$N$ junction in the absence of EF is written as
\begin{equation}
P(\epsilon_0,\Delta_R=0) =|A_{LR}|^2=\frac{t^2}{t^2+4\cos^2 [\Psi(\epsilon_0)]} .\label{transmission-Final}
\end{equation}
By making use of Landauer-B\"{u}ttiker approach \cite{SFEprb,quantuminterf,Moskalets} we obtain in the limit of a small transport voltage the linear ballistic conductance  as
\begin{equation}
    \sigma=4G_0\sum_{p_y}P(\epsilon_F),
    \label{Landauer}
\end{equation}
where $\epsilon_F$ is the Fermi energy, which can be tuned by an externally applied gate voltage $V_g$, and $G_0=\frac{2e^2}{h}$ is the quantum conductance. The numerical coefficient $4$ in the Eq. (\ref{Landauer}) is due to the presence of spin and valley degrees of freedom.
Since the electron transmission through  interfaces is small, i.e. $t \ll 1$, we obtain a small dc conductance $\sigma~\propto~t^2$ for arbitrary values of $V_g$. However, a strong enhancement of the total transmission can be obtained in $N$-$P$-$N$ junctions for particular values of gate voltages. This phenomenon is the resonant tunneling through the quasi-bound states determined by condition $\cos \Psi=0$ (see Eq. (\ref{transmission-Final})). In this case the ballistic conductance reaches the maximum value of $\sigma=4G_0$. The values of quasi-bound states can be found explicitly as
$\Psi_n(\epsilon_F)= \hbar \pi(n+1/2)$, or calculating the integral over $z$ in Eq. (\ref{Phase})
\begin{equation}
 a\sqrt{(U_0-\epsilon_F)_n^2-(\mathrm{v}p_y)^2}= \hbar \pi(n+1/2)  .
 \label{resonantvoltage}
\end{equation}

Since the parameter $U_0-\epsilon_F$ varies with the gate voltage, i.e.  $U_0-\epsilon_F~\propto~V_g$, the resonant tunneling appears as a set of symmetric resonant peaks in the dependence of total conductance $\sigma$ on the gate voltage $V_g$. The typical resonant peak in the dependence of $\sigma (V_g)$ is presented in Fig. 3 (the red (gray) curve). Similar resonances have been predicted and analyzed in graphene based quantum point contacts \cite{EfSilv}.
\begin{figure}[tbp]
\includegraphics[width=3in,angle=0]{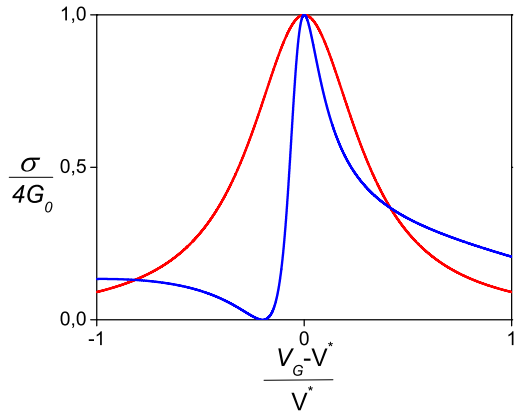}
\caption{The typical resonant peaks in the dependence of dc conductance on the gate voltage, $\sigma(V_g)$. The peaks shape changes from a symmetric one in the absence of EF (red (gray) curve) up to asymmetric one, i.e the Fano type, as the EF is applied [blue (dark) curve]. The voltage positions of resonances $V^*=U_0-\epsilon_F$ are determined by Eq. (\ref{resonantvoltage}) and Eq. (\ref{resonantvoltage-ph}).
} \label{Resonances}
\end{figure}

As the $N$-$P$-$N$ junction is irradiated by EF the phase trajectories become more complicated due to the presence of \emph{dynamical resonant points } located close to each interface [blue (dark) circles in Fig. 2b] . In the region closed to such points the resonant emission (absorption) of photons occurs.  The typical phase trajectories in this case are shown in Fig. 2b. Notice  here, that the electrons can move on the closed trajectory shown in Fig. 2b by blue (dark) line, in the presence of a standard Landau-Zener tunneling only.

The dynamic tunneling through resonant points provides two "short paths" for propagating electrons through the interface of a junction. These paths allow the quasi-particles propagate through junction without the Landau-Zener tunneling on interfaces. A one path occurs in the valence band and other one occurs  in the conduction band. The interference between these paths leads to the \emph{quantum interference effect }, i.e. the oscillations of transmission coefficient, considered previously in Ref. \cite{quantuminterf} for $N$-$P$ junction. In the case of  irradiated $N$-$P$-$N$ junctions we can write the transmission amplitude as
$$
A^{ph}_{LR}=4A_{\ell}B_{\ell}A_{r}B_{r}\cos \varphi_\ell \cos \varphi_r (1+X_{ph}+X_{ph}^2+....),
$$
$$
X_{ph}= (A_{\ell}^2 e^{-i\varphi_\ell}-B_{\ell}^2e^{i\varphi_\ell})(A_{r}^2 e^{-i\varphi_r}-B_{r}^2e^{i\varphi_r})
$$
\begin{equation}
\exp\{2i(\Psi_{ph}+\pi/2)\}\label{transmission-photon},
\end{equation}
where $A_{\ell (r)}$ and $B_{\ell (r)}$ are the amplitudes of dynamic transmission (reflection) on dynamical resonant points. These amplitudes are also satisfied to the relationship $A_{\ell (r)}^2+B_{\ell (r)}^2=1$. Similarly to a well-known analysis of Landau-Zener tunneling \cite{Tunn} the dynamic transmission coefficient $T_{\ell (r)}=A_{\ell (r)}^2$ is written explicitly as \cite{FEprl,SFEprb,quantuminterf}
\begin{equation}
T_{\ell(r)}=\exp \left[-\frac{\Delta_R^2\lambda_{\ell(r)}}{\mathrm{v}\hbar U_0}\right]~.
\label{transmission-dynamic}
\end{equation}
The quantum-mechanical phases $\varphi_{\ell(r)}=(4/\hbar)\int_{z_{res}}^{z_{\ell(r)}}p_z(z)dz$ are determined by confinement of quasi-particles between the dynamical resonant points and turning points.

The $X_{ph}$ is the amplitude of quasi-particle propagation along a closed trajectory (see the red (gray) line in Fig. 2b). Such an amplitude is determined by the total quantum-mechanical phase $\Psi_{ph} $ of quasi-particles emitting a photon and  propagating from left dynamical resonant point  up to right dynamical resonant point as $\Psi_{ph}=\Psi (\epsilon_0-\hbar \omega)$ [see, Eq. (\ref{Phase})]. Notice here, that Eq. (\ref{transmission-photon}) is valid in the limit $t_{\ell(r)}<<1$ as we neglected by direct propagation of quasi-particles through the interfaces, i.e. the internal closed trajectory (blue (dark) line in Fig. 2b).

Calculating a sum in Eq. (\ref{transmission-photon}) we obtain
\begin{equation}
A^{ph}_{LR}=\frac{4A_{\ell}B_{\ell}A_{r}B_{r}\cos \varphi_\ell \cos \varphi_r}{1-X}
\label{transmission-amplitude-photon}
\end{equation}
Next, we consider a symmetric $N$-$P$-$N$ junction irradiated by a weak electromagnetic field. In this case, $T_{\ell}=T_{r}=T$, $\varphi_{\ell}=\varphi_{r}=\varphi$ and, moreover, $(1-T)~\propto \Delta_R^2 \ll 1$.  The transmission coefficient taking into account the resonant interaction of propagating electrons with electromagnetic field is given by formula:

\begin{equation}
P_{ph}(\epsilon_0) =|A^{ph}_{LR}|^2=\frac{(1-T)^2  \cos^4 \varphi }{(1-T)^2\cos^4 \varphi+4\cos^2 \Psi_{ph}(\epsilon_0)},\label{transmission-Final-EF}
\end{equation}

The Eqs. (\ref{Landauer}) and (\ref{transmission-Final-EF})  determine a set of resonant peaks in the dependence of $\sigma_{ph}(V_g)$. The typical resonant peak displaying the Fano shape is shown in Fig. 3 (blue curve). The voltage positions of resonances in irradiated $N$-$P$-$N$ junctions are determined by conditions $\Psi_{ph}(\epsilon_F)=\hbar \pi(n+1/2)$ or explicitly:
\begin{equation}
 a\sqrt{[(U_0-\epsilon_F)_n+\hbar \omega]^2-(\mathrm{v}p_y)^2}= \hbar \pi(n+1/2)  .
 \label{resonantvoltage-ph}
\end{equation}

\section{Discussion and conclusions}

In conclusion we theoretically studied the quantum resonances in irradiated graphene based $N$-$P$-$N$ junctions. These quantum resonances manifest themselves by a set of sharp peaks in the dependence of dc ballistic conductance $\sigma_{ph}$ on the gate voltage $V_g$. The resonances show an asymmetric line shape similar to  the Fano resonances well-known in atomic physics \cite{Fano} and physics of nanostructures \cite{FanoCM,FanoFistul}. This peculiar form results from the combination of two effects: the resonant tunneling through microwave induced quasi-bound states, and the quantum interference effect between two microwave induced paths allowing quasi-particles penetrate and escape the $P$-region of a junction.  The typical resonance is shown in Fig. 3 (blue (dark) line). The maximum value of microwave induced dc conductance $\sigma_{ph}$ occurs for gate voltages determined by Eq. (\ref{resonantvoltage-ph}) and it shows a negative shift, $~\simeq~\hbar \omega$, in respect to the resonances obtained in non-radiated $N$-$P$-$N$ junctions. The drop values of resonances are determined by condition $\varphi_{\ell(r)}=\pi (n+1/2)$. The typical value of the microwave induced dc ballistic conductance, i.e. the value of $\sigma_{ph}$ measured far away from the resonances, strongly depends on the power of applied radiation $S$ as: $\sigma_{ph} \propto S^2$.

In order to observe these resonances some important conditions has to be satisfied. E.g.  since the direct tunneling spoils the microwave induced resonances, the transmission coefficient has to be small, i.e. $t_{\ell(r)}<<1$. Therefore, the values of transverse momentum $p_y \gg \sqrt{\hbar U_0/(\mathrm{v}\lambda)}$ have not be too small. Moreover, the resonances corresponding to the adjacent values of $p_y $ has to be well separated in order to do not wash out the resonances.  Thus, the $N$-$P$-$N$  junctions prepared in graphene nanoribbons with relatively small width $W$ are suitable to observe these microwave-induced resonances. Similar Fano-type resonances can be observed in N-P-N junctions prepared in bilayer graphene, semiconducting carbon nanotubes and /or graphene based quantum point contacts.

We acknowledge the financial support from the SPP 1459 "Graphene" and the SFB
Transregio 12 by DFG, and  the financial support of the Ministry of Education and Science of the Russian Federation  in the framework of Increase Competitiveness Program of NUST "MISiS"($K2-2014-015$).

\end{document}